\documentstyle[epsf,aps,prd,preprint,tighten,eqsecnum]{revtex}

\begin{document}
\draft

\preprint{\vtop{\hbox{RU99-9-B}\hbox{hep-ph/9908248}}}

\title{On the Perturbative Nature of Color Superconductivity}
\author{William E. Brown${}^a$, James T. Liu${}^{a,b}$
and Hai-cang Ren${}^{a,c}$}
\address{${}^a$ Department of Physics, The Rockefeller University,\\
 1230 York Avenue, New York, NY 10021.}
\address{${}^b$ Randall Laboratory of Physics, University of Michigan,\\
Ann Arbor, MI 48109.}
\address{${}^c$ Department of Natural Science, Baruch College of CUNY,\\
New York, NY 10010.}

\maketitle

\begin{abstract}
Color superconductivity is a possible phase of high
density QCD.  We present a systematic derivation of the transition
temperature, $T_C$, from the QCD Lagrangian through study of the
di-quark proper vertex.  With this approach, we confirm the dependence
of $T_C$ on the coupling $g$, namely $T_C\sim\mu g^{-5}e^{-\kappa/g}$,
previously obtained from the one-gluon exchange approximation in the
superconducting phase.  The diagrammatic approach we employ allows us
to examine the perturbative expansion of the vertex and the propagators.
We find an additional $O(1)$ contribution to the prefactor of the
exponential from the one-loop quark self energy and that the
other one-loop radiative contributions and the two gluon exchange vertex
contribution are subleading.
\end{abstract}


\widetext

\section{Introduction.}

Color superconductivity is a possible phase of high density QCD,
pioneered by Bailin and Love and others, \cite{bailin1984}, who
pointed out that the $\bar {\bf 3}$ channel of a di-quark interaction
is attractive through one gluon exchange.  Recently, using an effective
four-fermion interaction in the superconducting phase, much work has been
carried out on examining an energetically favored condensate which, for
$N_f=3$, breaks the original $SU(3)^c\times SU(3)^{f_L}\times SU(3)^{f_R}$
symmetry to its diagonal symmetry
\cite{alford1998a,rapp1998,alford1998b,schafer1998}.  This color
superconductivity mechanism has been called color-flavor locking.

Several attempts have been made to determine the parameters of the
superphase from the QCD action
\cite{pisarski1998,son1998,pisarski1999,schafer1999,%
pisarski1999b,pisarski1999c,hong1999}.
In all cases one arrives at the same dependence of the
zero-temperature gap energy on the QCD running coupling constant
evaluated at the Fermi energy, namely 
\begin{equation}
\Delta\sim \mu g^{-5}
e^{-{\kappa/g}},
\label{eq:nbcs}
\end{equation}
where $\mu$ is the chemical potential and $g$ the running Yang-Mills
coupling evaluated at $\mu$.
This dependence upon the coupling differs significantly
from the BCS case, $\Delta \sim \mu e^{-{\kappa^\prime/g^2}}$, due to the
long range propagation of magnetic gluons.  The coefficient
$\kappa=3\pi^2/\sqrt{2}$ was first derived by Son \cite{son1998}, and
subsequently verified in \cite{schafer1999} and
\cite{pisarski1999b,pisarski1999c}.
The latter two of these generalized this scaling behavior to the transition
temperature and in addition derived its ratio to the zero temperature
gap.  All of these results were obtained from superconducting gap
equations with one gluon exchange.  As such, although they contain the 
correct leading order behaviour that determines the dependence of the 
gap on the coupling, more detailed calculations are required to recover 
all the leading order contributions to the pre-exponential factor.  

In contrast with previous works, we approach the transition temperature
from the normal phase, with $T\ll \mu$. There are several 
advantages with this approach. First of all, the propagators in the normal 
phase are not subjected to modifications from the long range order, 
the form of which are Ansatz dependent. This ensures that issues of gauge
invariance and higher order corrections are relatively simple to handle.
Secondly, the integral equation for 
the proper vertex function, which determines the pairing instability, is 
linear while the gap equation in the superphase is nonlinear. Thirdly, the 
hard dense loop contribution to the gluon propagator is free from the 
Meissner effect. Though it has not been taken into account so far in the 
gap equation calculations in the superphase, the Meissner effect is
nevertheless expected to alter the pre-exponential factor \cite{schafer1999}.

Working directly from the QCD Lagrangian provides
a natural framework within which to examine the perturbative nature of
the theory at high density.  Indeed, we shall find that interactions
of second order, $O(g^4)$, make a leading order contribution to the
pre-exponential factor and that all higher order contributions are
subleading.  The sum of these effects suggests that a derivation of the
transition temperature from the normal phase will not only provide a
rigorous verification of the results obtained from within the
superphase, but should also allow a clean and exact determination of
the pre-exponential factor.  Combined with the aesthetic benefit of
calculating directly from the QCD Lagrangian, this more than
compensates for the technical complexity of this approach.

Starting from the $SU(N)$ QCD
Lagrangian, the calculation of the transition temperature can be cast
easily into thermal diagrams with gauge invariance manifest.
Following the formulation developed by Gor'kov and Melik-Barkhudarov
for the non-relativistic Fermi-gas, \cite{gorkov1961}, which allows
for a systematic examination of the higher order contributions, we
calculate the transition temperature to leading order in coupling and
obtain
\begin{equation}
\pi k_BT_C =  \mu g^{-5} c c' e^{-\sqrt{{6N\over N+1}}{\pi^2\over g}},
\label{eq:transt}
\end{equation}
where up to a constant of $O(1)$, $c=1024 \sqrt{2}\pi^4 N_f^{-5/2}$,
in agreement with \cite{schafer1999,pisarski1999b,pisarski1999c},
and the exponent, $-\sqrt{{6N\over N+1}}{\pi^2\over g}$, in agreement with
\cite{son1998,schafer1999,pisarski1999b,pisarski1999c}, are determined by the
leading order one gluon exchange process.  The previously unreported factor
\begin{eqnarray}
c'&=&\exp \left[{-\frac{1}{16} (\pi^2 + 4) (N-1)}\right]\nonumber\\
&\simeq&0.1766\qquad\hbox{for $N=3$},
\end{eqnarray}
comes from the
logarithmic suppression of the quasi-particle weight in the dressed quark
propagator.  For Landau damping, obtained in the hard dense loop
approximation, the contribution to the prefactor
from two gluon exchange diagrams is subleading in $g$.  This, however, is
not the case for a hypothetical static screening case where the perturbative
nature is completely spoiled by infrared log-enhancement in higher orders.

\section{Calculation of the QCD Transition Temperature.}

We consider an $SU(N)$ color gauge field coupled to $N_f$ flavors 
of massless quarks with the Lagrangian density
\begin{equation}
{\cal L}=-{1\over 4}F_{\mu\nu}^aF^a_{\mu\nu}-\bar\psi_f\gamma_\mu
(\partial_\mu-igA_\mu)\psi_f,
\label{eq:qcd}
\end{equation}
where $F_{\mu\nu}^a=\partial_\mu A_\nu^a-\partial_\nu
A_\mu^a+gf^{abc}A_\mu^bA_\nu^c$, $A_\mu=A_\mu^at^a$ with $t^a$ the
$SU(N)$ generator in its fundamental representation.  Since the
Lagrangian (\ref{eq:qcd}) is diagonal with respect to both flavor and
chirality, the corresponding indices will be dropped below.

\begin{figure}
\epsfxsize 6cm
\centerline{\epsffile{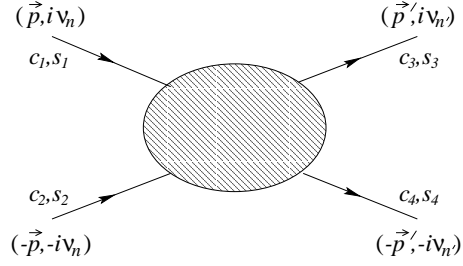}}
\bigskip
\caption{Proper vertex function,
$\Gamma_{s_3,s_4;s_1,s_2}^{c_3,c_4;c_1,c_2}(n',n|\vec p\,',\vec p\,)$.}
\label{fig1}
\end{figure}

\begin{figure}
\epsfxsize 8cm
\centerline{\epsffile{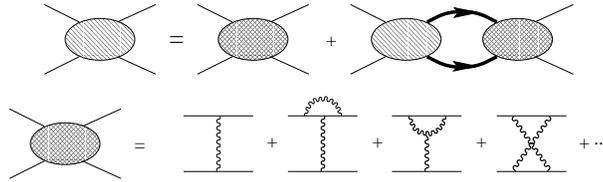}}
\bigskip
\caption{The Schwinger-Dyson equation.  As in Fig.~\ref{fig1},
$\Gamma$ is represented by singly hashed vertices and $\tilde\Gamma$
is represented by double hashed vertices.  The expansion of $\tilde\Gamma$
is given up to $O(g^4)$.}
\label{fig2}
\end{figure}

We derive the transition temperature by investigating the onset of the
pairing instability in the proper vertex function corresponding to the
scattering of two quarks at non-zero temperature and chemical potential.
This vertex function with zero total momentum and zero total
Matsubara energy, 
$\Gamma_{s_3,s_4;s_1,s_2}^{c_3,c_4;c_1,c_2}(n',n|\vec p\, ',\vec p\,)$,
is shown in Fig.~\ref{fig1}, where $n$ and $n'$ label
the Matsubara energies $i\nu_n={2\pi i\over\beta} (n+{1\over 2})$ of 
individual quarks.
Each of the superscripts $c$, which denote color, are associated with
a leg.  The subscripts $s$, which label the states above or below the
Dirac sea, are either $+$ or $-$.  We find it convenient for the
partial wave analysis to associate the Dirac spinors $u(\vec p\,)$ and
$v(\vec p\,)$, which satisfy the Dirac equations
$(\gamma_4p-i\vec\gamma\cdot\vec p\,)u(\vec p\,)=0$ and
$(\gamma_4p-i\vec\gamma\cdot\vec p\,)v(\vec p\,)=0$, to the
quark-gluon vertex instead of to the quark propagator. Therefore, where $s=(s_1,s_2)$ represents the incoming subscripts and $s'=(s_3,s_4)$ represents the outgoing subscripts,
suppressing the color indices and momentum-energy dependence, we write
\begin{equation}
\Gamma_{s';s} = \bar U_{\gamma}(s_3,\vec p\,') \bar
U_\delta(s_4,-\vec p\,')\Gamma_{\gamma\delta,\alpha\beta}
U_{\alpha}(s_1,\vec p\,)U_{\beta}(s_2,-\vec p\,).
\end{equation}
The vertex function $\Gamma_{\gamma\delta,\alpha\beta}$ is given by
conventional Feynman rules, and $U(s,\vec p\,)=u(\vec p\,)$ or $v(-\vec p\,)$
for $s=+$ or $-$, respectively.  The proper vertex function satisfies a
Dyson-Schwinger equation as shown in Fig.~\ref{fig2}.  This integral
equation of Fredholm type may be written with all indices suppressed as
\begin{equation}
\label{eq:ie}
\Gamma(n^\prime,n\vert\vec p\,',\vec p\,)
=\tilde\Gamma(n^\prime,n\vert\vec p\,',\vec p\,)+
{1\over\beta}\sum_m\int{d^3\vec q \over (2\pi)^3}
K(n^\prime,m\vert \vec p\,',\vec q\,)
\Gamma(m,n\vert\vec q,\vec p\,),
\end{equation}
where $\tilde\Gamma$ represents the two quark irreducible vertex with
all indices defined in the same way as for $\Gamma$.  The kernel has the 
explicit form
\begin{equation}
K_{s_3,s_4;s_1,s_2}^{c_3,c_4;c_1,c_2} (n^\prime,m\vert \vec p\,',\vec
q\,) = \tilde\Gamma_{s_3,s_4;s_1,s_2}^{c_3,c_4;c_1,c_2}
(n^\prime,m\vert\vec p\,',\vec q\,) S_{s_1}(m\vert\vec
q\,)S_{s_2}(-m\vert-\vec q\,),
\end{equation}
where we write $S_s(n\vert\vec p\,)$
for the full quark propagator with momentum $\vec p$ and Matsubara
energy $i\nu_n$.  The zeroth order quark propagator reads
\begin{equation}
S_s(n\vert\vec p\,)={i\over i\nu_n-sp+\mu},
\end{equation}
and the diagrammatic expansion of
$\tilde\Gamma$ to $O(g^4)$ is displayed in
Fig.~\ref{fig2}.

\begin{figure}
\epsfxsize 8cm
\centerline{\epsffile{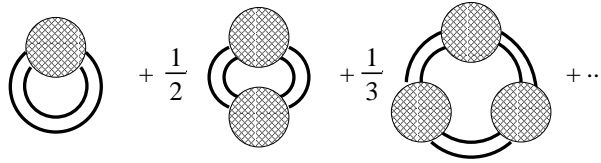}}
\bigskip
\caption{The Bubble Diagrams.}
\label{fig3}
\end{figure}

The transition temperature may be obtained from (\ref{eq:ie}) by
examining the Fredholm determinant, ${\cal D}\equiv \det(1-K)$, which is a
function of $T$ and $\mu$.  $T_C$ is then given by the highest temperature
at which ${\cal D}(T,\mu)$ vanishes. To demonstrate the gauge invariance 
of this formulation, we need to extend the integral equation (\ref{eq:ie})
to include
scattering with arbitrary total momentum and energy. Denote by $K^\prime$ the 
kernel for which either the total momentum or the total energy or both are 
nonzero. Then the Fredholm determinant with arbitrary total momentum and energy 
factorizes as
\begin{equation}
\hat{\cal D}=\det(1-K)\det(1-K^\prime).
\end{equation}
On the other hand, 
$\ln{\hat{\cal D}}$ is given by the sum of bubble diagrams shown in
Fig.~\ref{fig3}. They are in fact manifestly gauge invariant as may be seen
following the argument given in \cite{thooft}.

For the rest of this paper, we shall work in Coulomb gauge, in which the 
full gluon propagator takes the form
\begin{eqnarray}
\label{eq:glue1}
D_{ij}(\vec k , \omega)&=&D^M(\vec k, \omega)
(\delta_{ij}-{k_ik_j\over \vec k^2}),\\
\label{eq:glue2}
D_{44}(\vec k , \omega)&=&D^E(\vec k , \omega),\\
\noalign{and}
\label{eq:glue3}
D_{4j}(\vec k , \omega)&=&D_{j4}(\vec k , \omega)=0.
\end{eqnarray}

Since $\Gamma$ corresponds to di-quark scattering it can be decomposed into
irreducible representations of $SU(N)$ by
either symmetrization [representation {\bf{6}} for $SU(3)$] or
antisymmetrization [representation $\bar {\bf 3}$ for $SU(3)$] among the
initial and final color indices, {\it i.e.}
\begin{eqnarray}
\Gamma_{s^\prime,s}^{c^\prime, c}(n^\prime,n\vert\vec p\,',\vec p\,)
&=& \sqrt{2}\;\delta^{c_1(c_3}\delta^{c_4)c_2}
\Gamma_{s^\prime,s}^S(n^\prime,n\vert\vec p\,', \vec p\,) \\ \nonumber
&+& \sqrt{2}\;\delta^{c_1[c_3}\delta^{c_4]c_2}
\Gamma_{s^\prime,s}^A(n^\prime,n\vert\vec p\,', \vec p\,),
\label{eq:symanti}
\end{eqnarray}
where $(\cdots)$ and $[\cdots]$ denote symmetrization and
antisymmetrization with weight one, respectively.  $\tilde\Gamma$ may be
decomposed similarly.  Since the Fermi surface has a pairing
instability in the presence of even an arbitrarily weak attractive
interaction, as is also the case for BCS theory, we need only focus on
the attractive antisymmetric channel for non-Abelian theories.

Both $\tilde\Gamma_{s^\prime,s}^A(n^\prime,n\vert\vec p\,',\vec p\,)$ and
$\Gamma_{s^\prime,s}^A(n^\prime,n\vert\vec p\,',\vec p\,)$ can be
expanded in terms of Legendre polynomials, {\it i.e.}
\begin{equation}
\Gamma_{s^\prime,s}^A(n^\prime,n\vert\vec p\,',\vec p\,)
=\sum_l\gamma_{s^\prime,s}^l(n^\prime,n\vert p^\prime,p)
P_l(\cos\theta).
\label{eq:legend}
\end{equation}
Substituting  such expansions into (\ref{eq:ie}), we find another Fredholm
equation satisfied by $\gamma_{s^\prime,s}^l(n^\prime,n\vert p^\prime ,p)$:
\begin{equation}
\label{eq:7}
\gamma_{s^\prime,s}^l(n^\prime,n\vert p^\prime,p)
=\tilde\gamma_{s^\prime,s}^l (n^\prime,n\vert p^\prime,p)
+{1\over\beta}\sum_{m,s^{\prime\prime}}
\int_0^\infty dq\,K_{s^\prime,s^{\prime\prime}}^l(n^\prime,m\vert p^\prime,q)
\gamma_{s^{\prime\prime},s}^l(m,n\vert q,p),
\end{equation}
where the kernel $K_{s',s}^l$ has the form 
\begin{equation}
K_{s^\prime,s}^l(n^\prime,n\vert p^\prime,p)
={p^2\tilde\gamma_{s^\prime,s}^l(n^\prime,n\vert p^\prime,p)
\over 2\pi^2(2l+1)}
S_{s_1}(n\vert p)
S_{s_2}(-n\vert p).
\label{kernel}
\end{equation}
We consider the $l=0$ term in the partial wave expansion, although
the higher partial wave terms may contribute \cite{schafer1999}.

The Fredholm determinant of (\ref{eq:7}) with $l=0$ can be written as
${\cal D}=\prod_j (1-\lambda_j^{-2})$ with $\lambda_j^2$ the eigenvalues
(labeled by the integer suffix $j$) defined by the homogeneous
equation,
\begin{equation}
f_{s'}(n,p)=\lambda^2{1\over\beta}\sum_{m,s}
\int_0^\infty dq\,K_{s^\prime,s}^0(n,m\vert p,q)
f_{s}(m,q).
\label{eq:hint}
\end{equation}
At sufficiently high temperature, all $\lambda_j^2 > 1$, so that
${\cal D}\ne0$ and there is no instability---the theory is in the normal
phase.  As the temperature is reduced, we find the transition
temperature to the superconducting phase is that at which the smallest
of $\{\lambda_j^2\}$ reaches one.  The solution of (\ref{eq:hint})
provides the eigenvalues in terms of the parameters of the theory;
the temperature, coupling and chemical potential.  Hence the
inversion of $\lambda_0^2(T,g,\mu)=1$, where $\lambda_0$ is the
smallest eigenvalue, yields the transition temperature $T_c$.

In the presence of a Fermi sea, hard dense loops have to be included in
the gluon propagator. As a result, the Coulomb interaction is strongly
screened by the Debye length, $\lambda_D = m_D^{-1}$, where
\begin{equation}
m_D^2={N_f g^2\over \pi^2}
\int_0^\infty dq\,q\,{1\over e^{\beta(q-\mu)}+1}\simeq \frac{N_f g^2
\mu^2}{2 \pi^2}.
\end{equation}
The dressed Coulomb propagator at momentum-energy
$(\vec k, i \omega)$ reads 
\begin{equation}
\label{eq:cp}
D^E(\vec k ,\omega) = \frac{-i}{\vec k^2 + \sigma^E(\vec k, \omega)},
\end{equation}
with $\sigma^E(\vec k,\omega) \simeq m_D^2$ for $\omega \ll k \ll \mu$.
However, the magnetic interaction is poorly screened. While a magnetic
mass may exist, of order $T$, Landau damping \cite{Bellac} prevails at
$\mu\gg k_BT$.
In this case the propagator for a magnetic gluon is  
\begin{equation}
D^M(\vec k,\omega)={-i\over \vec k^2+\omega^2+\sigma^M(\vec k,\omega)}.
\label{eq:glue}
\end{equation}
The only region of infrared sensitivity on the $(k,\omega)$-plane is
$\omega \ll k\ll \mu$ where $\sigma^M(\vec k, \omega)
\simeq{\pi\over 4}m_D^2\frac{|\omega|}{|\vec k|}$. To $g^2$ order,
the contribution to $\tilde{\gamma}_{++}^0(n^\prime,n\vert
p^\prime,p)$ arises from the one-gluon exchange diagram of Fig.~\ref{fig2} and
is given by
\begin{equation}
\label{eq:gam}
\tilde\gamma_{++}^0(n^\prime,n\vert p^\prime,p) = -{g^2\over 12p^\prime p}
\left(1+{1\over N}\right)
\left[\ln{8\mu^3\over|p^\prime-p|^3+{\pi\over 4}m_D^2|\nu_{n^\prime}
-\nu_n|} + \frac{3}{2} \ln \frac{4 \mu^2}{m_D^2} \right],
\end{equation}
where a term finite in the limit $g \rightarrow 0$ has been dropped.
To the leading order of $\ln\mu/ k_BT$, the summation over Matsubara
energy can be replaced by an integral and $|p' - p|^3$ ignored.
Furthermore, the components $f_{+-}$, $f_{-+}$ and $f_{--}$ can be
neglected and the integral over
$q$ can be carried out for a solution smooth in the neighborhood of the
Fermi sea, $p\simeq \mu$.
Equation (\ref{eq:hint}) is then approximated by
\begin{equation}
f(\nu) = \int_\epsilon^{\nu_0} {d\hat\nu^\prime\over\hat\nu^\prime} 
{\cal K}(\nu,\nu^\prime)f(\nu^\prime),
\label{eq:inteq}
\end{equation}
where
\begin{equation}
\hat\nu = {N_f^{5/2} g^5 \over1024 \sqrt{2} \pi^4 \mu}\nu,
\end{equation}
and
\begin{equation}
\epsilon = {N_f^{5/2} g^5 k_BT\over1024 \sqrt{2} \pi^3 \mu}.
\end{equation}
The reduced kernel ${\cal K}(\nu,\nu')$ is given by 
\begin{equation}
\label{eq:redk}
{\cal K}(\nu,\nu^\prime) = {\lambda^2\over 2}{g^2\over 24\pi^2}
\left(1+{1\over N}\right)
\left(\ln{1\over |\hat\nu-\hat\nu^\prime|}
     +\ln{1\over |\hat\nu+\hat\nu^\prime|}\right),
\end{equation}
with an ultraviolet cutoff $\nu_0\sim 1$ introduced.  The
eigenvalue problem (\ref{eq:inteq}) can be solved using the same
approximation employed by Son \cite{son1998}, which amounts to replacing
the kernel of (\ref{eq:inteq}) with $(2/\hat\nu') \ln 1/\hat\nu_>$ where
$\hat\nu_>= \max(\hat\nu,\hat\nu^\prime)$.  We find that
\begin{equation}
\label{eq:evals}
\lambda_j^2 \frac{g^2}{24\pi^2} \left(1 + \frac{1}{N}\right) \ln^2
\frac{1}{\epsilon} = \left( j + \frac{1}{2} \right)^2 \pi^2,
\end{equation}
where $j$ is an integer and
\begin{equation}
f(\nu) \simeq \sqrt{\frac{2}{\ln 1/\epsilon}}
\sin\left[\left(j + \frac{1}{2}\right) \pi \frac{\ln 1/\hat\nu}{\ln 1/\epsilon}
\right].
\end{equation}
Setting the smallest eigenvalue to one, namely $\lambda_0^2 = 1$, we finally
arrive at the exponent and the prefactor $c$ of (\ref{eq:transt}).

Here we wish to highlight the mathematical structure of (\ref{eq:ie}) which
characterizes the long range attractive interaction.  If, instead, the pairing
interaction was of a short range nature, the transition temperature could be
located by means of the standard expansion of the Fredholm determinant
\begin{eqnarray}
{\cal D}(\lambda)=1&-&{1\over\beta}\sum_n\int_0^\infty \!\!dp\,K(n,p|n,p)
\nonumber\\
&+&{1\over 2\beta^2}\sum_{n,n^\prime}\int_0^\infty \!\!dp\int_0^\infty \!\!dq
\left|\matrix{\tabskip12pt K(n,p|n,p)&K(n,p|n',q)\cr K(n',q|n,p)&K(n',q|n',q)}
\right|+\cdots.
\label{eq:fd}
\end{eqnarray}
The $m^{\rm{th}}$ term in the expansion on the right hand side contains $m$
kernels folded together and for a short range interaction would be 
of the order $g^{2m}\ln{1\over\epsilon}$ since there is only one eigenvalue
of $K$ which diverges as $\ln{1\over\epsilon}$ in the limit $\epsilon\to0$.
At the transition temperature one would have
$g^2\ln{1\over\epsilon}\sim 1$, and thus the order of magnitude of the
subsequent terms would be $g^{2(m-1)}$. Therefore only the first two
non-trivial terms ($m=1,2$) would be sufficient to determine the transition
temperature up to the leading order of the pre-exponential factor. On the
other hand, for the present long range attraction of the QCD model, the
logarithm in the kernel of (\ref{eq:hint}) makes the $m^{\rm{th}}$ term of
the expansion (\ref{eq:fd}) of the order $g^{2m}\ln^{2m}{1\over\epsilon}$
since there are now an infinite number of eigenvalues of $K$ which diverge
as $\ln^2{1\over\epsilon}$ in the limit $\epsilon\to0$, as indicated in
(\ref{eq:evals}).
Hence the series can not be truncated at $T_C$, and a new method has to be
employed to fix both the exponent and the prefactor.

Note that the second term on the right hand side of (\ref{eq:fd}) is in fact
logarithmically
infinite for the kernel (\ref{kernel}).  On the other hand, this term
corresponds to the sum $\sum\lambda_j^{-2}$ which is convergent 
according to (\ref{eq:evals}). The reason for this apparent paradox lies in
Son's approximation, which led to the eigenvalues given in (\ref{eq:evals}).
When the approximation is corrected, the eigenvalues instead become
\begin{equation}
\lambda_j^{-2}={g^2\over24\pi^4}\left(1+{1\over N}\right)
\left[\left(j+{1\over 2}\right)^{-2}\ln^2
{1\over\epsilon}+c_j\right],
\end{equation}
with $c_j\sim j^{-1}$ for $j\gg1$ \cite{blrnext}, which 
explains the appearance of the logarithmic divergence in this term when
summing over $j$.

It is interesting to note that a similar
non-BCS scaling behavior of the type indicated in (\ref{eq:nbcs})
was obtained via a mean field calculation of 2D superconductivity at the
von~Hove singularity of the electronic density of states \cite{tsuei}.

\section{Higher Order Corrections.}

We now come to the question of higher order corrections to the kernel
from the perturbative expansion of the quark propagator and the
irreducible vertex $\tilde\Gamma$.  These become important if
sufficient powers of the infrared logarithm accompany the coupling
constant $g^2$.  It follows from (\ref{eq:transt}) that
$\ln{\mu/k_BT_C}\sim1/g$ at the transition temperature. Thus, if the
$O(g^4)$ contribution to $K_{s',s}^0$ is of the form
$g^4\ln^\delta\mu/|\nu_{n^\prime}-\nu_n|$ or
$g^4\ln^\delta\mu/|\nu_{n^\prime}|$, its magnitude relative to the $O(g^2)$
term given by
(\ref{eq:gam}) will be $g^{3-\delta}$.  The perturbative robustness of
the exponent then requires that $\delta<3$; robustness of the
prefactor requires $\delta<2$.  We find with Landau damping that
$\delta=2$; one logarithm originates from the leading order of
$\tilde{\gamma}$, (\ref{eq:gam}), and the second from the self-energy
of quarks \cite{son1998}.  Therefore there will be an $O(1)$ correction to
the prefactor.

Parametrizing the dressed quark propagator above the Dirac sea as
\begin{equation}
S_+(p_0,\vec p)={i\over p_0-p+\mu-\Sigma(p_0,\vec p)},
\end{equation}
we find, to one loop order, that
\begin{equation}
\label{eq:16}
\left. \frac{\partial}{\partial p_0} {\rm Re}\, \Sigma(p_0, \vec p )   
\right|_{p=\mu} = -\frac{N^2 -1}{N} \frac{g^2}{24\pi^2} \ln
\frac{\mu^3}{m_D^2 \max(|p_0|,k_BT)},
\end{equation}
while $\left. \frac{\partial}{\partial p}{\rm Re}\, \Sigma(p_0, \vec p )
\right|_{p=\mu}$ remains finite in the limit $p_0\to\infty$. For a Matsubara
energy $p_0=i\nu_n=i{2\pi\over\beta}(n+{1\over 2})$, we thus have
\begin{equation}
S_+(n|p)={i\over
i\left(1+{N^2-1\over N}{g^2\over24\pi^2}\ln{\mu^3\over m_D^2 |\nu_n|}\right)
\nu_n-p+\mu}.
\end{equation}
Following the steps which lead from (\ref{eq:cp}) to (\ref{eq:redk}) above,
the inclusion of this effect amounts to replacing the reduced kernel 
${\cal K}(\nu,\nu^\prime)$ by ${\cal K}(\nu,\nu^\prime)+\Delta
{\cal K}(\nu,\nu^\prime)$ where
\begin{equation}
\Delta{\cal K}(\nu,\nu^\prime)=-\frac{N^2 -1}{N} \frac{g^2}{24\pi^2} 
{\cal K}(\nu,\nu^\prime)\ln \frac{1}{\hat\nu'}.
\end{equation}
Treating $\Delta{\cal K}(\nu,\nu^\prime)$ as a perturbation, the shift of 
the eigenvalue in (\ref{eq:evals}) with $j=0$ turns out to be
\begin{eqnarray}
\delta{1\over\lambda_0^2}&=&\int_\epsilon^{\nu_0}{d\hat\nu\over\hat\nu}
\int_\epsilon^{\nu_0}{d\hat\nu^\prime\over\hat\nu^\prime}f_0(\nu)\Delta{\cal K}
(\nu,\nu^\prime)f_0(\nu^\prime)\nonumber\\
&=&-{2(\pi^2+4)\over\pi^4}\left(1+{1\over N}\right){N^2-1\over N}
\left({g^2\over24\pi^2}\right)^2\ln^3{1\over\epsilon}.
\end{eqnarray}
The condition for the critical temperature, $\lambda_0^2=1$, now becomes
\begin{equation}
{g^2\over6\pi^4}\left(1+{1\over N}\right)\ln^2{1\over\epsilon}
-{2\over\pi^4}\left({g^2\over24\pi^2}\right)^2\left(1+{1\over N}\right)
{N^2-1\over N}(\pi^2+4)\ln^3{1\over\epsilon}=1,
\end{equation}
the solution of which, for small $g$, gives rise to the result
(\ref{eq:transt}) with both $c$ and $c'$.  The
logarithmic dependence of (\ref{eq:16}) upon the coupling constant $g$,
written in $m_D^2$, will change the prefactor $g^{-5}$ of (\ref{eq:transt})
to $g^{-5+O(g)}$.  However this correction is of higher order.

Other higher order corrections to $\tilde{\gamma}$ have also been partially
addressed in the literature. The vertex correction has been discussed
in \cite{schafer1999} and some renormalization group arguments have been
applied to the straight box diagrams of two gluon exchange,
\cite{son1998}, which was conjectured to be of the same order as the
crossed box diagram. For the case of Landau damping, our analysis of
the vertex correction is in agreement with \cite{schafer1999}, indicating an
$O(g)$ contribution to the prefactor.  We also find that the
crossed box diagram is free from any logarithmic enhancement and so
its contribution is suppressed relative to the one gluon exchange by a
factor of $g^2/\ln (\mu/k_B T)$ \cite{prprivate}. On the other hand,
the straight box
diagram, which corresponds to the convolution of two single gluon
exchange kernels, is logarithmically divergent at $T=0$ and contains
all the powers of logarithms at $T\ne0$ necessary to produce the result of the
ladder sum implicit in (\ref{eq:hint}).

The crossed box contributes to $\tilde{\gamma}_{++}^0$ a term 
\begin{eqnarray}
B &=& -{1\over 2}g^4\int_{-1}^1d\cos\theta\int_{-\infty}^\infty {d\omega
\over 2\pi}\int{d^3\vec l\over(2\pi)^3}
D_{\mu\mu^\prime}(\vec l-{\vec q\over 2},i\omega)
D_{\nu\nu^\prime}(\vec l+{\vec q\over 2},i\omega)\nonumber\\
&&\times [\bar u(\vec p\,')\gamma_\mu S_F(\vec P+\vec l,i\omega)
\gamma_\nu u(\vec p\,)]
[\bar u(-\vec p\,')\gamma_{\nu^\prime}S_F(-\vec P+
\vec l,i\omega)\gamma_{\mu^\prime}u(-\vec p\,)],
\end{eqnarray}
where $|\vec p\,|=|\vec p\,'|=\mu$, $\vec P={1\over 2} (\vec p+\vec
p\,')$, $\vec q=\vec p-\vec p\,'$ and the summation over Matsubara
energy has been replaced by an integral for $T \ll \mu$.  Ignoring the
Coulomb propagator, the contribution from the small scattering angle,
$|\theta| < \theta_0\ll 1$,
and infrared region, ${\rm IR:}\,|\vec l\,| \ll \mu$ and $|\omega| \ll \mu$, 
\begin{eqnarray}
B_{\rm{IR}} &=& -{1\over 2}g^4\int_{-\theta_0}^{\theta_0}d\theta\sin\theta
\int_{\rm IR} {d\omega\over 2\pi}\,{d^3\vec l\over(2\pi)^3}
D_{ii^\prime}(\vec l- {\vec q\over 2},i\omega)
D_{jj^\prime}(\vec l+{\vec q\over 2},i\omega)\nonumber\\
&&\times[\bar u(\vec p\,')\gamma_i S_F(\vec P+\vec l,i\omega)
\gamma_j u(\vec p\,)]
[\bar u(-\vec p\,')\gamma_{j^\prime}S_F(-\vec P+
\vec l,i\omega)\gamma_{i^\prime}u(-\vec p\,)],
\end{eqnarray}
is bounded: $B_{\rm IR}\leq b$ where
\begin{equation}
b \equiv {1\over 32\pi^4\mu^2}\int_0^{\theta_0}d\theta\int_{\rm IR}d\rho\,
d^3\vec r
{r_+r_-|E_+E_--\rho^2|\over(r_+^3+\kappa|\rho|)(r_-^3+\kappa|\rho|)(\rho^2+
E_+^2)(\rho^2+E_-^2)}.
\end{equation}
Here $E_\pm=|\vec P\pm\vec l\,|/\mu-1$, 
$r_\pm=|\vec l\pm\vec q\,|/\mu$, $\rho = |\omega|/\mu$ and
$\kappa={\pi\over4}{m_D^2\over\mu^2}$.
Transforming the integration variables 
from $\theta,\vec r$ to $E_\pm,r_\pm$, we end up with
$b={1\over 32\pi^4\mu^2}\int_0d\rho K(\rho)$ where
\begin{equation}
K(\rho)=
\int dE_+dE_-dr_+^2dr_-^2\,J
{r_+r_-|E_+E_--\rho^2|\over(r_+^3+\kappa|\rho|)
(r_-^3+\kappa|\rho|)(\rho^2+E_+^2)(\rho^2+E_-^2)},
\end{equation}
with the Jacobian 
\begin{equation}
J=[(E_+-E_-)^4-4(r_+^2+r_-^2)(E_+-E_-)^2-16(E_++E_-)^2+16r_+^2r_-^2]^{-{1/2}}.
\end{equation}
As $\rho\to 0$, we find that $K(\rho)\to{\rm{const~}\times}
\rho^{-{2/3}}$ up to some power of $\ln\rho$. Therefore $B_{IR}$ as 
well as $B$ is free from infrared divergences.

For the sake of comparison, we have also examined the $O(g^4)$ corrections 
with only a static mass gap for gluons, $m\ll\mu$. This amounts to replacing
the magnetic gluon propagator (\ref{eq:glue}) by  
\begin{equation}
D^M(\vec k,\omega)={-i\over \omega^2+\vec k^2+m^2}.
\end{equation}
The integration over 
the Euclidean energy can be carried out readily by the residue theorem and 
the remaining integral can be classified according to the contributions from
the gluon and quark poles. Denoting the $O(g^4)$ 
contribution to $\tilde{\gamma}_{++}^0$ by $\Delta
\tilde\gamma$, we find to the leading order in $\ln{\mu/ m}$ 
\begin{equation}
\Delta\tilde\gamma={g^4\over 2\pi^2}
\left[{1\over 2N}\left(1+{1\over N}\right) c_1
-{1\over 4}\left(N-{2\over N}-{1\over N^2}\right)c_2\right]
\ln^3{\mu\over m},
\end{equation}
where $c_1$ corresponds to the vertex corrections and $c_2$
corresponds to the crossed box diagram in Fig.~\ref{fig2}. We write
$c_1=g_1+q_1$ and $c_2=g_2+q_2$, where $g_1$ and $g_2$ come from the
gluon poles and $q_1$ and $q_2$ come from the quark poles, respectively.
The $g$ and $q$ coefficients are tabulated in Table~\ref{tbl:llc} for various
cases. This presence of $\ln^3\mu/m$ will ruin the perturbative nature of the
formulation, provided $m\sim T$.
At this point, the difference between Landau damping and static
screening is clear. To the leading order of one gluon exchange, the
difference merely amounts to the substitution of the static screening
mass by $m_D^{2/3}|\omega|^{1/3}$, with $\omega$ the Euclidean energy
transfer and $|\omega|\ll\mu$. This is not at all the case
for higher order corrections, including two gluon exchange, even
though the infra-red sensitive region for the loop momentum of the
quark pole contribution in the case of static screening coincides with
that of Landau damping.

\section{Conclusion.}

In conclusion, we have derived the superconducting transition
temperature with thermal diagrams in the normal phase.  This ensures
gauge invariance to all orders.  We have also examined systematically
the $O(g^4)$ corrections and found an additional contribution to the
pre-exponential factor of $T_c$ in the literature.  Unlike the situations
with static screening, Landau damping significantly improves the infra-red
behavior of the higher order diagrams and makes the perturbative
expansion of the prefactor in terms of $g$ legitimate.

In a forthcoming publication \cite{blrnext}, we develop a perturbative 
formulation that enables us to eliminate the $O(1)$ uncertainty of 
the prefactor $c$ of Eqn.~(\ref{eq:nbcs}).
Since the perturbative expansion is given in terms of $g$ or $g\ln g$,
the application to realistic situations is in fact rather limited.
As was pointed out in \cite{schafer1999}, even with asymptotic freedom,
$g=0.67$ at $\mu=10^{10}$Mev, and yet the energy scale probed at RHIC is
only a few hundred Mev.  Nevertheless, it reveals some novel
properties of superconductivity induced by a long range interaction
which has not yet been fully examined in the literature.

\bigskip
\noindent
{\bf Acknowledgements.}

We are grateful to R. Pisarski and D. Rischke
for stimulating discussions and valuable suggestions at various stages
of this work. Research supported in part by the U.S. Department of Energy under grant DOE-91ER40651-TASKB.

\newpage

\begin{table}
\begin{tabular}{c|c|c|c|c}
&\omit\span transverse\kern24pt&\omit\span covariant\kern24pt\\
\hline
&with FS&without FS&with FS&without FS\\
\hline
$g_1$&$-1/4$&$-1/4$&0&0\\
$q_1$&0&0&0&0\\
$g_2$&0&0&$-2/3$&$-2/3$\\
$q_2$&$1/16$&0&$1$&0
\end{tabular}
\bigskip
\caption{Leading log coefficients}
\label{tbl:llc}
\end{table}

\end{document}